\documentclass[prd,twocolumn,superscriptaddress,preprintnumbers,nofootinbib]{revtex4}
\usepackage{graphicx}
\usepackage{epsfig}
\usepackage{bm}
\usepackage{latexsym,amssymb,amsmath,amsfonts,amssymb,wasysym,float}
\usepackage{scrextend}
\usepackage{mathrsfs}
\usepackage{color}

\usepackage{enumitem}

\newcommand{\postscript}[2]{\setlength{\epsfxsize}{#2\hsize}
   \centerline{\epsfbox{#1}}}

\usepackage[usenames,dvipsnames]{xcolor}
\definecolor{orange}{cmyk}{0,0.5,1,0}
\definecolor{rossoCP3}{cmyk}{0,.88,.77,.40}
\definecolor{graa}{rgb}{0.8,0.8,0.8}
\definecolor{blaa}{rgb}{0.2,0.2,0.6}
\usepackage{xspace}

\begin{document}
\preprint{MPP-2024-91}
\preprint{LMU-ASC 05/24}

\title{\color{rossoCP3} From infinite to infinitesimal: Using the
  Universe as a dataset to probe Casimir  corrections to the vacuum energy from fields inhabiting the dark dimension}

\author{Luis A. Anchordoqui}

\affiliation{Department of Physics and Astronomy,  Lehman College, City University of
  New York, NY 10468, USA
}

\affiliation{Department of Physics,
 Graduate Center, City University
  of New York,  NY 10016, USA
}

\affiliation{Department of Astrophysics,
 American Museum of Natural History, NY
 10024, USA
}

\author{Ignatios Antoniadis}

\affiliation{High Energy Physics Research Unit, Faculty of Science, Chulalongkorn University, Bangkok 1030, Thailand}

\affiliation{Laboratoire de Physique Th\'eorique et Hautes \'Energies
  - LPTHE, Sorbonne Universit\'e, CNRS, 4 Place Jussieu, 75005 Paris, France
}

\author{Dieter\nolinebreak~L\"ust}

\affiliation{Max--Planck--Institut f\"ur Physik,  
 Werner--Heisenberg--Institut,
80805 M\"unchen, Germany
}

\affiliation{Arnold Sommerfeld Center for Theoretical Physics, 
Ludwig-Maximilians-Universit\"at M\"unchen,
80333 M\"unchen, Germany
}

\author{Neena T. Noble}

\affiliation{Department of Physics and Astronomy,  Lehman College, City University of
  New York, NY 10468, USA
}

\affiliation{Department of Astrophysics,
 American Museum of Natural History, NY
 10024, USA
}

\author{Jorge F. Soriano}
\affiliation{Department of Physics and Astronomy,  Lehman College, City University of
  New York, NY 10468, USA
}

\begin{abstract}
  \vskip 2mm \noindent Promptly after high-resolution experiments
  harbinger the field of precision cosmology low- and high-redshift
  observations abruptly gave rise to a tension in the measurement of
  the present-day expansion rate of the Universe ($H_0$) and the
  clustering of matter ($S_8$). The statistically significant
  discrepancies between the locally measured values of $H_0$ and $S_8$
  and the ones inferred from observations of the cosmic microwave
  background assuming the canonical $\Lambda$ cold dark matter (CDM)
  cosmological model have become a new cornerstone of theoretical
  physics. $\Lambda_s$CDM is one of the many beyond Standard Model
  setups that have been proposed to simultaneously resolve the 
  cosmological tensions. This setup relies on an empirical conjecture,
  which postulates that $\Lambda$ switched sign (from
  negative to positive) at a critical redshift $z_c \sim 2$. We
  reexamine a stringy model that can describe the transition in the vacuum
  energy hypothesized in $\Lambda_s$CDM.
 The model makes use of the Casimir forces driven by fields
 inhabiting the incredible bulk of the dark dimension scenario. Unlike
the $\Lambda_s$CDM setup the model deviates from $\Lambda$CDM in the
early universe due to the existence of relativistic neutrino-like
species. Using the Boltzmann solver {\tt CLASS} in combination with
{\tt MontePython} we confront predictions of the stringy model to experimental data (from the
Planck mission, Pantheon+ supernova type Ia, BAO, and KiDS-1000). We show that
the string-inspired model provides a satisfactory fit to the data and
can resolve the cosmological tensions.
\end{abstract}
\date{April 2024}

\maketitle 

\section{Introduction}

Just about a century after the expansion of the Universe was
established~\cite{Hubble:1929ig}, the Hubble constant, which measures
its rate ($H_0 \equiv 100 h~{\rm km/s/Mpc}$), continues to encounter
challenging shortcomings. Inferring $H_0$ from the measured spectra of
temperature and polarization anisotropies in the cosmic microwave
background (CMB) requires a cosmological model to describe the
evolution of the baryon-photon fluid at early times (i.e., before the
last-scattering surface) as well as the evolution of the Universe at
later times (i.e., after the last-scattering surface). $\Lambda$ cold
dark matter (CDM) is the simplest model that can provide a good
phenomenological fit to current cosmological data~\cite{ParticleDataGroup:2022pth}, and therefore it is
generally adopted as benchmark to determine $H_0$. Likewise,
$\Lambda$CDM is assumed in the CMB determination of the growth of cosmic structure (parameterized by
$S_8 \equiv \sigma_8 \, \sqrt{\Omega_m/0.3}$, where $\sigma_8$ describes the matter fluctuations at
scales of $8~{\rm Mpc}/h$ and $\Omega_m$ is the present day value of
the non-relativistic matter density).

The $h = 0.674 \pm 0.005$ and
$S_8= 0.834 \pm 0.016$ inferred from {\it Planck}'s CMB data assuming
$\Lambda$CDM~\cite{Planck:2018vyg} are in $\sim 5\sigma$ tension with
$h =0.73 \pm 0.01$ from the SH0ES distance ladder
measurement (using Cepheid-calibrated type-Ia
supernovae)~\cite{Riess:2021jrx,Murakami:2023xuy} and in
$\sim 3\sigma$ tension with $S_8= 0.766^{+0.020}_{-0.014}$ from the
cosmic shear data of the Kilo-Degree Survey~\cite{Heymans:2020gsg},
respectively. As a matter of fact, it has been suggested that the
$H_0$ tension is actually a tension on the type Ia supernovae absolute magnitude $M_B$~\cite{Efstathiou:2021ocp,Camarena:2021jlr}, because the SH0ES $H_0$ measurement comes directly from $M_B$
estimates. Counting them all, these statistically significant discrepancies have
become a new cornerstone of theoretical physics, and many beyond
Standard Model (SM) physics models are rising to the
challenge~\cite{DiValentino:2021izs,Schoneberg:2021qvd,Perivolaropoulos:2021jda,Abdalla:2022yfr}.

Of particular interest herein, $\Lambda_s$CDM is one of the models that have
been proposed to simultaneously resolve the $H_0$, $S_8$, and $M_B$ tensions~\cite{Akarsu:2019hmw, Akarsu:2021fol, Akarsu:2022typ,Akarsu:2023mfb}. This model relies on an empirical conjecture which postulates
that $\Lambda$ switched sign (from negative to positive) at
critical redshift $z_c \sim 2$;
 \begin{equation} \Lambda\quad\rightarrow\quad\Lambda_{\rm s}\equiv \Lambda_0 \ {\rm sgn}[z_c-z],
\label{Lambdas}
 \end{equation}
 where $\Lambda_0>0$ and ${\rm sgn}[x]=-1,0,1$ for $x<0$, $x=0$
 and $x>0$, respectively. Apart from resolving the three major cosmological
 tensions, $\Lambda_s$CDM achieves quite a good fit to Lyman-$\alpha$
 data provided $z_c \alt 2.3$~\cite{Akarsu:2019hmw}, and it is in agreement with the otherwise puzzling JWST
observations~\cite{Adil:2023ara,Menci:2024rbq}.\footnote{An alternative model that accommodates the data has been presented in~\cite{Gomez-Valent:2024tdb}.}
 
Despite the remarkable success of $\Lambda_s$CDM to accommodate current experimental
data, the model is theoretically
unsatisfactory because it postulates that
the Universe experienced a rapid transition
from an anti-de Sitter (AdS) vacuum to a de Sitter (dS) vacuum, and
this hods out against the AdS swampland distance conjecture, which
posits that at zero temperature AdS and dS vacua are an infinite distance appart in
metric space~\cite{Lust:2019zwm}. In this work, we reexamine a stringy model
developed elsewhere~\cite{Anchordoqui:2023woo}, which can justify the
AdS $\to$ dS transition hypothesized in $\Lambda_s$CDM. Bearing this
in mind, throughout we
will refer to the model as $\Lambda_s$CDM$^{+}$.

The layout of the paper is as follows. In Sec.~\ref{sec:2} we first review 
the basic setting of $\Lambda_s$CDM$^{+}$ and after that we discuss phenomenological
implications of the model. In
Sec.~\ref{sec:3} we carry out a numerical analysis to confront
$\Lambda$CDM$^+$ with current astrophysical and cosmological
observations. The paper wraps up in Sec.~\ref{sec:4} with 
conclusions and some discussion.

\section{Swamplandish Cosmology}
\label{sec:2}

\subsection{The Dark Dimension}

Low-energy effective field theories (EFTs) are the driving gear in the
description of low-energy particle physics, cosmology, and
gravitational phenomena. It is common ground that the SM and General
Relativity should both be understood as leading terms in an EFT
expansion. Fundamental principles of string theory or black hole
physics can enforce consistency requirements on the space of
consistent low-energy EFTs. Indeed, this is the goal of the Swampland
program in the quest to understand which are the ``good''
low-energy EFTs that can couple to gravity consistently (e.g. the
landscape of superstring theory vacua) and distinguish them from the
``bad'' ones that cannot~\cite{Vafa:2005ui}. In theory space, the
frontier discerning the good theories from those downgraded to the
swampland is drawn by a family of conjectures classifying the
properties that an EFT should call for/avoid to enable a consistent
completion into quantum gravity~\cite{Palti:2019pca,vanBeest:2021lhn,Grana:2021zvf,Agmon:2022thq,Antoniadis:2024sfa}.

For example, the distance conjecture states that infinite
distance limits $\Delta \phi \to \infty$ in the moduli space of
massless scalar fields are accompanied by an infinite tower of
exponentially light states $m \sim e^{-\alpha \Delta \phi}$, where
distance and masses are measured in Planck
units~\cite{Ooguri:2006in}. Connected to the distance conjecture is
the AdS distance conjecture, which correlates the dark energy density
to the mass scale $m$ characterizing the infinite tower of states, $m
\sim |\Lambda|^\alpha$, as the negative AdS vacuum energy $\Lambda \to
0$, where $\alpha$ is an ${\cal O}(1)$ positive constant~\cite{Lust:2019zwm}. In
addition, under the hypothesis that this scaling behavior holds in dS
(or quasi dS) space, an unbounded number of massless modes also pop up
in the limit $\Lambda \to 0$.

The AdS distance conjecture in dS 
space provides a pathway, dubbed the dark dimension
scenario~\cite{Montero:2022prj}, to clear up the origin of the cosmological hierarchy $\Lambda/M_p^{4} \sim 10^{-120}$, because it connects the
size of the compact space $R_\perp$ to the
dark energy scale $\Lambda^{-1/4}$ via
\begin{equation}
  R_\perp \sim \lambda \ \Lambda^{-1/4} \,,
\label{RperpLambda}  
\end{equation}  
where $M_p$ is the reduced Planck mass and the proportionality factor is estimated to be within the range
$10^{-1} < \lambda < 10^{-4}$.\footnote{Auger data of highest energy cosmic rays favor $\lambda
  \sim 10^{-3}$~\cite{Anchordoqui:2022ejw,Noble:2023mfw}.} Actually, (\ref{RperpLambda}) derives from
  constraints by theory and experiment. On the theoretical side, since the Kaluza-Klein (KK)  tower contains massive spin-2 bosons, the Higuchi
  bound~\cite{Higuchi:1986py} places an absolute upper limit to the exponent of $\Lambda^\alpha$,
 whereas explicit string calculations
of the vacuum energy~(see
e.g.~\cite{Itoyama:1986ei,Itoyama:1987rc,Antoniadis:1991kh,Bonnefoy:2018tcp})
set a lower bound on $\alpha$.\footnote{The unitarity bound of Higuchi
  states that any massive spin-2 particle in dS space should be
  heavier than the expansion rate $H$~\cite{Higuchi:1986py}.} Altogether, these constraints yield $1/4 \leq \alpha \leq 1/2$. On the experimental side,
 constraints on deviations from Newton's
gravitational inverse-square law~\cite{Lee:2020zjt} and neutron star
heating~\cite{Hannestad:2003yd} give rise to the conclusion encapsulated
in (\ref{RperpLambda}): {\it The cosmological hierarchy problem can be
  addressed if there exists one extra dimension of
radius $R_\perp$ in the micron range, and the lower bound for $\alpha =
1/4$ is basically saturated~\cite{Montero:2022prj}.}  Within this
picture, the SM must be localized on a D-brane transverse to the dark dimension.
 A theoretical
amendment on the connection between the cosmological and KK mass scales confirms $\alpha =
1/4$~\cite{Anchordoqui:2023laz}.

A point worth noting at this juncture is that the KK tower of the dark 
dimension opens up at the mass scale $m_{\rm KK} \sim
1/R_\perp$ in the eV range. As a consequence, in the dark dimension scenario the five-dimensional (5D)  Planck scale (or species scale
where gravity becomes
strong~\cite{Dvali:2007hz,Dvali:2007wp,Cribiori:2022nke,vandeHeisteeg:2023dlw})
is found to be
\begin{equation}
M_* \sim m_{\rm KK}^{1/3} \ 
M_p^{2/3}   \sim 10^9~{\rm GeV} \, .
\label{species_f}
\end{equation}
Alternatively, it was recently speculated that the dark dimension can
be understood as a line interval with end-of-the-world 9-branes
attached at each end. Of course this is equivalent to a semicircular
dimension endowed with $S^1/\mathbb{Z}_2$ symmetry~\cite{Schwarz:2024tet}.

Adding to the story, the dark dimension scenario has the potential to explain the
smallness of neutrino masses. This proposal envisions the
right-handed neutrinos as 5D bulk states with
Yukawa couplings to the left-handed lepton and Higgs doublets that are localized states on the
SM
brane~\cite{Dienes:1998sb,Arkani-Hamed:1998wuz,Dvali:1999cn}. The neutrino masses are then suppressed due to the
wave function of the bulk states. To be
more specific, this is accomplished by introducing three 5D Dirac
fermions  $\Psi_\alpha$, which are singlets under the SM gauge
symmetries and interact in our brane with the three active left-handed
neutrinos in a way that conserves lepton number.  The  $S^1/\mathbb{Z}_2$ symmetry in the 5th dimension coordinate $x_5$
 contains $x_5$ to $-x_5$, which acts as chirality ($\gamma_5$) on
spinors. Then, in the Weyl basis each Dirac field can be decomposed into two two-component spinors $\Psi_\alpha \equiv (\psi_{\alpha L},\psi_{\alpha R})^T$.

The generation of neutrino masses
originates in 5D bulk-brane interactions of the form
\begin{equation} 
  \mathscr{L}  \supset h_{ij} \ \overline L_i \ \tilde{H} \ \psi_{jR}(x_5=0) \,,
\end{equation}
where $\tilde{H} = -i\sigma_{2}H^{*}$, $L_i$ denotes the lepton
doublets (localized on the SM brane), $\psi_{jR}$ stands for the three bulk (right-handed) $R$-neutrinos
evaluated at the position of the SM brane, $x_5=0$ in the
fifth-dimension, and $h_{ij}$ are coupling
constants. This gives a coupling with the
$L$-neutrinos of the form $\langle H \rangle \  \overline{\nu}_{L_i} \
\psi_{jR} (x_5=0)$, where $\langle H \rangle = 175~{\rm GeV}$ is the Higgs vacuum
expectation value. Expanding $\psi_{jR}$ into modes canonically normalized leads for each of them to a Yukawa $3 \times 3$ matrix suppressed by the square root of the volume of the bulk
$\sqrt{\pi R_\perp M_s}$, i.e.,
\begin{equation}
Y_{ij}= \frac{h_{ij}}{\sqrt{\pi R_\perp M_s}} \sim h_{ij} \frac{M_s}{M_p} \,,
\end{equation}
where $M_s \lesssim M_*$ is the string scale, and where
in the second rendition we have dropped factors of $\pi$'s and of the string coupling.

Now, neutrino oscillation data can be well-fitted in terms of two
nonzero differences $\Delta m^2_{ij} = m^2_i - m^2_j$ between the
squares of the masses of the three mass eigenstates; namely,
$\Delta m_{21}^2 =(7.53 \pm 0.18) \times 10^{-5}~{\rm eV}^2$ and
$\Delta m^2_{32} = (2.453 \pm 0.033) \times 10^{-3}~{\rm eV}^2$ or
$\Delta m^2_{32} = -(2.536 \pm 0.034) \times 10^{-3}~{\rm
  eV}^2$~\cite{ParticleDataGroup:2022pth}. It is easily seen that to
obtain the correct order of magnitude of neutrino masses the coupling
$h_{ij}$ should be of order $10^{-4}$ to $10^{-5}$ for
$10^9 \lesssim M_s/{\rm GeV} \lesssim
10^{10}$~\cite{Anchordoqui:2022svl}.  In the presence of bulk masses~\cite{Lukas:2000wn,Lukas:2000rg},
the mixing of the first KK modes to active neutrinos can be
suppressed~\cite{Carena:2017qhd,Anchordoqui:2023wkm}.

The latest chapter in the story contemplates dark matter
candidates. The dark dimension 
provides a colosseum for dark matter contenders. In particular, it was observed in~\cite{Gonzalo:2022jac,Law-Smith:2023czn,Obied:2023clp} that the universal coupling of
the SM fields to the massive spin-2 KK excitations of the graviton in
the dark dimension provides a dark matter candidate. The ``dark-to-dark''
decays driving the dynamics of the KK modes provide a specific realization of dynamical dark matter framework~\cite{Dienes:2011ja}. Complementary to
the dark gravitons, it was discussed in~\cite{Anchordoqui:2022txe,Anchordoqui:2024akj,Anchordoqui:2024dxu,Anchordoqui:2024jkn} that primordial black
holes with Schwarzschild radius smaller than a micron could also be
good dark matter candidates, possibly even with an interesting close
relation to the dark gravitons~\cite{Anchordoqui:2022tgp}. Finally, the dark dimension can also
accommodate fuzzy dark matter~\cite{Anchordoqui:2023tln}.

All in all, the dark dimension facilitates a structured scenario to
describe the cosmological evolution of the dark sector. We now turn to
discuss how this scenario could also help to resolve the current discordances
in the cosmological parameters.

Before proceeding, we pause to note that the idea of
  considering large extra dimensions
to bring down $M_p$ (or else
  $M_s$) down to near the electroweak scale was proposed in~\cite{Arkani-Hamed:1998sfv,Antoniadis:1998ig,Arkani-Hamed:1998jmv}. It is
  also worthy to mention that the motivation for considering such a large compact space is rather
  different from the rationale behind the dark dimension scenario. Namely,
  {\it large extra dimensions were introduced with the aim of
resolving the electroweak hierarchy problem through unification of weak scale with a higher dimensional Planck scale in the TeV
range, whereas the idea encapsulated in (\ref{species_f}) ties the existence of a mesoscopic extra dimension to the cosmological hierarchy problem rather than the electroweak one.}

\subsection{$\bm{\Lambda_s}$CDM$^{\bm{+}}$}

Inspired by the discussion in the previous section, we consider 5D Einstein-de Sitter gravity compactified on a circle $S^1$ endowed
with $S^1/\mathbb{Z}_2$ symmetry, and we assume that the SM is
localized on a D-brane transverse to the compact 5th dimension, whereas gravity spills into the dark dimension. The 
effective 4D potential of the modulus controlling the
radius (or radion field) $R$ is found to be  
\begin{equation}
  V(R) = \frac{2 \pi \ \Lambda_5 \ r^2}{R} +\left(\frac{r}{R} \right)^2 \ T_4  + V_C (R) \,,
\label{V}
\end{equation}
where $\Lambda_5$ is the 5D cosmological constant, $r \equiv \langle
R \rangle$ is the vacuum expectation value of the radion, $T_4$ is the
total 3-brane tension of the model at the SM position, and $V_C$ stands for the quantum corrections to the vacuum energy due to Casimir forces~\cite{Anchordoqui:2023etp}. These corrections are expected to become important in the
deep infrared region, because the Casimir contribution to the potential falls off exponentially at
large $R$ compared to the particle wavelength. Indeed,  as $R$ decreases different particle thresholds open up, 
\begin{equation}
V_C (R) =  \sum_i \frac{\pi r^2}{32
   \pi^7 R^6} \ (N_F - N_B) \ \Theta (R_i -R) \,,
 \label{xxx}
\end{equation} 
where $m_i =
 R_i^{-1}$ are the masses of the 5D fields, $\Theta$ is a step function, and $N_F - N_B$ stands for the
 difference between the number of light fermionic and bosonic degrees
 of freedom. At the classical level, i.e. considering
 only the first two terms in (\ref{V}), it is straightforward to see that the
 potential develops a maximum at
\begin{equation}
 R_{\rm max} = - T_4/(\pi
 \Lambda_5) \,,
\end{equation}
requiring a negative tension $T_4$. The stringy origin of
  the negative nature of $T_4$ is discussed in the Appendix. Now, note that if the fermionic degrees of freedom overwhelm the bosonic contribution, they would give rise to possible
minima, as long as $R_i < R_{\rm max}$.
\begin{figure}[htb!]
    \postscript{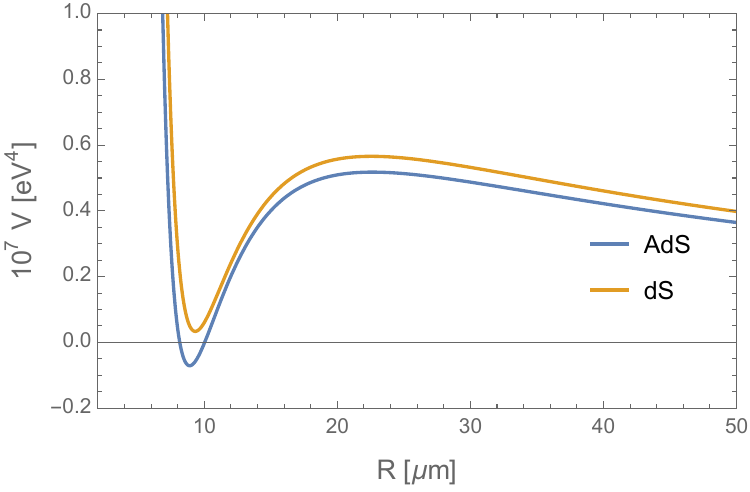}{0.9}
  \caption{The potential $V(R)$  for $(\Lambda_5)^{1/5} =
    22.6~{\rm meV}$ and $|T_4|^{1/4} = 24.2~{\rm meV}$ considering $N_F-N_B = 6$
    (AdS) and $N_F-N_B = 7$ (dS). 
   \label{fig:1}}
\end{figure}

We now have to define the 5D mass spectrum. Counting the six 5D Weyl
fields we find that neutrinos contribute with $N_F = 12$. On the other
hand, the 5D graviton contributes with 5 bosonic degrees of
freedom. In addition, we assume that the  5D spectrum contains a real
scalar field $\varphi$. We further assume that $\varphi$ has a
potential with two local minima, with very small difference in vacuum
energy and bigger curvature (mass) of the lower one. Around $z\sim 2$
the false vacuum  ``tunnels'' to its true vacuum state. After the
quantum tunneling $\varphi$ becomes more massive and its contribution
to the Casimir energy becomes exponentially suppressed. Altogether,
this implies that for $z \gtrsim 2$, the number of bosonic degrees of
freedom is $N_B = 6$, but for $z \lesssim 2$, it reduces to $N_B = 5$.
In Fig.~\ref{fig:1} we show an illustrative example of the phase
transition of AdS $\to$ dS vacua generated by the dark sector described above. 

Our analysis has three caveats which are worthy of mention:
\begin{itemize}[noitemsep,topsep=0pt]
\item The 5D vacuum transition creates a $\delta V$ contribution to
$\Lambda_5$, where $\delta V = V_{\rm min}- V_{\rm localmin}$ corresponding to the vacuum energies of the upper (localmin) and the
lower (min) minima. We take  $\delta V \ll \Lambda_5$ so that it does
not perturb the analysis producing the curves shown if
Fig.~\ref{fig:1}.
\item All terms of the potential that
stabilize the radius are of the same order. So there is no fine tuning
among them for the vacuum energy, but there is a hierarchy of all
these scales compared to the fundamental scale which has to be
assumed.
However, this assumption is part of the distance conjecture for the
cosmological constant~\cite{Lust:2019zwm}.
\item The Higuchi bound~\cite{Higuchi:1986py} translates into a necessary
  condition for the consistency of the effective theory described by (\ref{V}),
  \begin{equation}
    (\Lambda_5/M_*^3)^{-1/2} > R \, ,
\end{equation}    
which implies a very low scale of
4D inflation if $R \sim 1~\mu{\rm m}$~\cite{Dvali:1998pa}. To accommodate this hierarchy we adopt the working assumption
introduced
elsewhere~\cite{Anchordoqui:2022svl,Anchordoqui:2023etp,Antoniadis:2023sya}
that the Universe undergoes a period of inflation in which the radius
of the dark dimension expanded exponentially fast, from the species
length up to the micron-scale. In such a uniform 5D inflation the
Higuchi bound is satisfied trivially before inflation (as the compact space is
of order $M_s^{-1}$), as well as at the end of inflation in terms of 4D quantities for any $\Lambda_5^{\rm end}$
less than $\Lambda_5^{\rm initial}$. 
\end{itemize}

Now, the AdS $\to$ dS transition shown in Fig.~\ref{fig:1} slightly deviates from the model
analyzed in~\cite{Akarsu:2023mfb}, because the fields characterizing the deep
infrared region of the dark sector contribute to the effective number
of relativistic neutrino-like
species $N_{\rm eff}$~\cite{Steigman:1977kc}. Using conservation of entropy, fully
thermalized relics with $g_*$ degrees of freedom contribute
\begin{equation}
  \Delta N_{\rm eff} = g_* \left(\frac{43}{ 4g_s}\right)^{4/3} \left
    \{ \begin{array}{ll} 4/7 & {\rm for  \ bosons}\\ 1/2 & {\rm for \
                                                           fermions} \end{array}
               \right.                                        \,,
\end{equation}
where $g_s$ denotes the effective 
degrees of freedom for the entropy of the other thermalized
relativistic species that are present when they decouple~\cite{Anchordoqui:2019amx}. The 5D
graviton has 5 helicities, but the spin-1 helicities do not have zero
modes, because we assume the compactification has
$S^1/\mathbb{Z}_2$ symmetry and so the $\pm 1$ helicities are
projected out. The spin-0 is the
radion and
the spin-2 helicities form the massless (zero mode) graviton. This means
that for the 5D graviton, $g_*=3$. The scalar field $\varphi$ contributes with $g_*=1$. The (bulk) left-handed neutrinos are odd, but the right-handed neutrinos are even and so each
counts as a Weyl neutrino, for a total $g_* =2 \times 3$. Assuming that the
dark sector decouples from the SM sector before the electroweak phase
transition we have $g_s = 106.75$. This gives $\Delta N_{\rm eff} =
0.25$. It is of interest to investigate whether these extra-relativistic
degrees of freedom can spoil the $\Lambda_s$CDM predictions for $H_0$
and $S_8$.

\section{Numerical Analysis}
\label{sec:3}

To implement the data analysis we use the Markov Chain Monte Carlo
technique. A Markov chain consists of a sequence of random numbers
called states. Each number is stochastically obtained from the
previous number without explicitly being dependent on it. In our
analysis the states of a chain consist on specific values of the free
parameters in a given cosmological model.

\begin{figure*}[htb!]
    \postscript{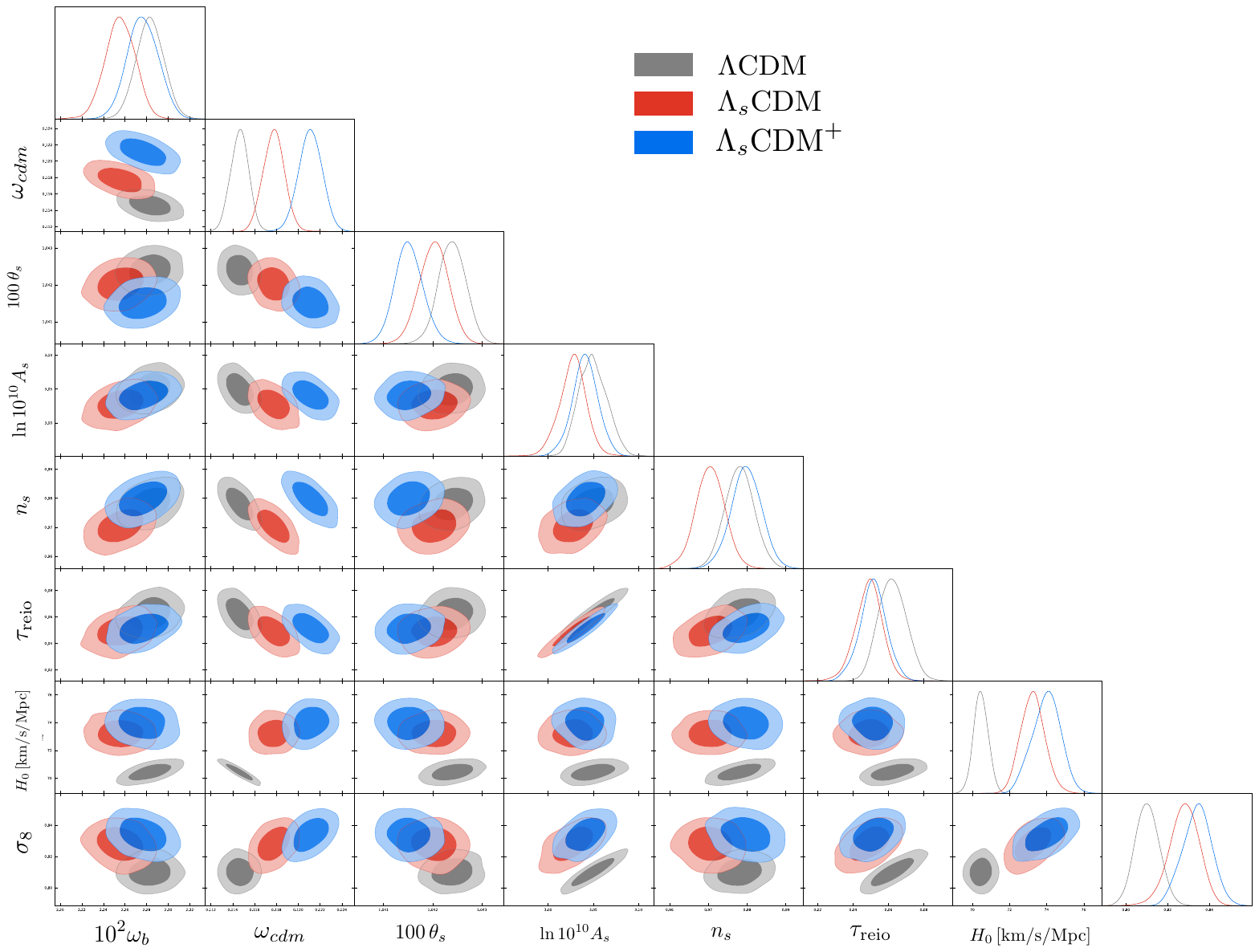}{0.95}
    \caption{One-dimensional posterior distributions and joint marginalized
      contours (at 68\% and 95\% CL) 
      of the free parameters of the $\Lambda$CDM,
      $\Lambda_s$CDM, and $\Lambda_s$CDM$^+$ models using data from
    {\it Planck}, transversal BAO, Pantheon+, and KiDS-1000.
   \label{fig:2}}
\end{figure*}

\begin{figure}[htb!]
    \postscript{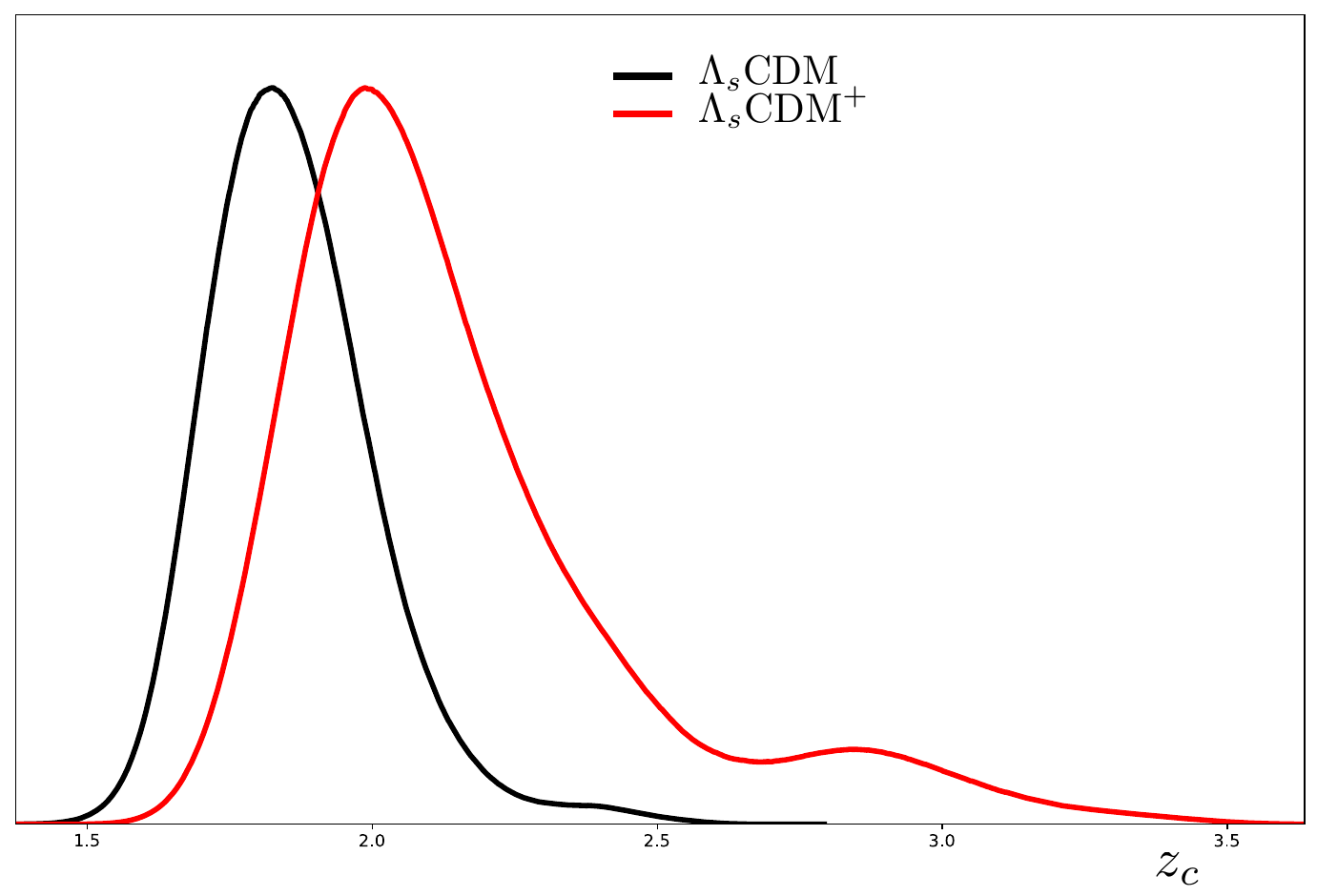}{0.95}
    \caption{One-dimensional 
      distribution of the critical redshift for the $\Lambda_s$CDM and $\Lambda_s$CDM$^+$ models using data from
      {\it Planck}, transversal BAO, Pantheon+, and KiDS-1000.
   \label{fig:3}}
\end{figure}

We adopt the {\tt CLASS+MontePython} code~\cite{Blas:2011rf,Lesgourgues:2011re, Brinckmann:2018cvx} to explore the full parameter space of the $\Lambda_s$CDM$^+$ model
and compare its predictions to $\Lambda$CDM and $\Lambda_s$CDM. We
derive constraints on the cosmological parameters of these three models from
 data-sets and likelihoods given below using 
the Metropolis-Hastings algorithm~\cite{Robert:2016}, while ensuring a Gelman-Rubin
convergence criterion of $R - 1 < 10^{-2}$ in all the
runs~\cite{Gelman:1992zz}. The baseline seven free parameters of the $\Lambda_s$CDM and
$\Lambda_s$CDM$^+$ models are given by ${\cal P} = \{\omega_b,\, \omega_{cdm},\,
  \theta_s, \, \tau_{\rm reio}, \,  n_s, \, A_s, \, z_c\}$, where the first six are the
common ones with $\Lambda$CDM:
{\it (i)}~the baryon density $\omega_b
\equiv \Omega_b h^2$, {\it (ii)}~the CDM density $\omega_{cdm} 
\equiv \Omega_c h^2$, {\it (iii)}~the angular size of the sound
horizon at recombination $\theta_s$, {\it (iv)}~the Thomson scattering
optical depth due to reionization $\tau_{\rm reio}$, {\it (v)}~the scalar
spectral index $n_s$, and {\it (vi)}~the power spectrum amplitude of adiabatic
scalar perturbations $A_s$. For $\Lambda$CDM and $\Lambda_s$CDM, we take
$N_{\rm eff} = 3.044$~\cite{Gariazzo:2019gyi}, whereas for $\Lambda_s$CDM$^+$, we take $N_{\rm
  eff} = 3.294$.

To constrain the free parameters of the  models we make use of a
series of astrophysical and cosmological probes:
\begin{itemize}[noitemsep,topsep=0pt]
\item \textbf{Planck 2018 CMB data}: The CMB temperature,
  polarization, and lensing angular power spectra from the {\it Planck} 2018 legacy release~\cite{Planck:2018vyg,Planck:2019nip}. 
\item \textbf{Transversal BAO}: Measurements of 2D baryon acoustic
  oscillations (BAO),
 $\theta_{\text{BAO}}(z)$, obtained in a weakly model-dependent
 approach, and compiled in Table~I of~\cite{Nunes:2020hzy}.
\item \textbf{Pantheon+}: The 1701 light curves of 1550 distinct
  supernovae type Ia, which are distributed in the redshift interval
  \mbox{$0.001 \leq z \leq 2.26$}~\cite{Brout:2022vxf}. We incorporate
  the most recent SH0ES Cepheid host distance anchors~\cite{Riess:2021jrx} into the
  likelihood function by integrating distance modulus measurements of
  the Pantheon+ supernovae.
\item \textbf{Cosmic Shear}: KiDS-1000
  data~\cite{Kuijken:2019gsa,Giblin:2020quj}, including the weak
  lensing two-point statistics data for both the auto and
  cross-correlations across five tomographic redshift
  bins~\cite{Hildebrandt:2020rno}. We follow the KiDS team analysis
  and adopt the COSEBIs (Complete Orthogonal Sets of E/B-Integrals)
  likelihood~\cite{KiDS:2020suj}. Nonlinearities are implemented into the
analysis using {\tt
  HALOFIT}~\cite{Smith:2002dz,Bird:2011rb}.\footnote{This
  implementation comes up with a
  difference to the data analysis carried out in~\cite{Akarsu:2023mfb}
  (which was executed using the  
 {\tt HMcode}~\cite{Mead:2015yca}), but such a difference leads to
 (almost) negligible effects in the results.}
\end{itemize}

\begin{table*}
 \caption{Marginalized constraints, mean values with 68\% CL, on 
   free parameters of the models. $\chi^2_{\rm min}$ is given in the last row.
\label{tabla1}}
\centering
\begin{tabular}{cccc}
  \hline
  \hline
    parameter & $\Lambda\mathrm{CDM}$ & $\Lambda_s\mathrm{CDM}$ & $\Lambda_s\mathrm{CDM}^+$\\\hline
~~~~~~$\omega_b$~~~~~~        & ~~~~~~$0.02283\pm0.000127$~~~~~~  &
                                                    ~~~~~~$0.022562\pm0.000139$~~~~~~
                                                                &
                                                                  ~~~~~~$0.022756\pm0.000139$~~~~~~ \\
$\omega_{cdm}$    & $0.114628\pm0.000815$ &$0.117709 \pm 0.000974$     &$0.12117\pm0.00106$\\
$100 \ \theta_s $    & $1.04239\pm0.000278$  &$1.04204\pm0.000298$    &$1.04151\pm0.00028$\\
$\ln(10^{10}A_s)$ & $3.0496\pm0.0154$    &$3.0289\pm0.0147$ 	     &$3.0400\pm0.0148$\\
$n_s         $    & $0.97806\pm0.00369$  &$0.97074\pm0.00381$    &$0.97978\pm0.00391$\\
$\tau_{\rm reio} $    & $0.06226\pm0.00784$  &$0.04958\pm0.00763$   &$0.05084\pm0.00769$\\
$z_c$             & -
                                      &$1.864\pm0.149$
                                                                &$2.105\pm0.264$\\
\hline
  $\chi_{\rm min}^2$ & $4224.7$ & $4192.64$ & $4195.94$ \\
  \hline
  \hline
    \end{tabular}
    \end{table*}

\begin{table*}\centering
\caption{Marginalized constraints, mean values with 68\% CL, on 
   derived parameters of the models. \label{tabla2}}
  \begin{tabular}{cccc}
  \hline
  \hline
    parameter & $\Lambda\mathrm{CDM}$ & $\Lambda_s\mathrm{CDM}$ & $\Lambda_s\mathrm{CDM}^+$ \\\hline
~~~~~~~~$H_0/ {\rm km/s/Mpc}$ ~~~~~~~~ & ~~~~~~~~$70.44\pm 0.39$~~~~~~~~
&~~~~~~~~$73.25\pm0.65$~~~~~~~~  & ~~~~~~~~$74.04\pm0.71$~~~~~~~~ \\
$\sigma_8$& $0.8102\pm0.0058$ &$0.8284\pm0.0070$&$0.8343\pm0.0071$ \\
  \hline \hline
    \end{tabular}
     \end{table*}

In Fig.~\ref{fig:2} we show the one-dimensional posterior
distributions and joint marginalized contours (at 68\% and
95\% CL)  of free and derived parameters of the
$\Lambda$CDM, $\Lambda_s$CDM, and $\Lambda_s$CDM$^+$ models.  The
corresponding marginalized 68\% CL
constraints on the baseline free parameters and selected derived parameters are listed in
Tables~\ref{tabla1} and \ref{tabla2}. One can immediately appreciate the qualitative similarities in the predictions of $\Lambda_s$CDM and $\Lambda_s$CDM$^+$. The best
parameter to distinguish the models seem to be $\omega_{cdm}$ and
$n_s$ has the potential to discriminate between $\Lambda_s$CDM and $\Lambda_s$CDM$^+$. There is quite a good
agreement between the  $\Lambda$CDM$^+$ predicted value of the Hubble
constant $h =
0.7404 \pm 0.0071$ and the SH0ES measurement
$h =0.73 \pm 0.01$~\cite{Riess:2021jrx,Murakami:2023xuy}. The
situation is even more interesting for $S_8$: an estimate of structure
growth predicted by $\Lambda_s$CDM$^+$ follows from  $\sigma_8 =
0.8343 \pm 0.0071$,  $\omega_b = 0.022756 \pm 0.000139$, $\omega_{cdm} =
0.12117 \pm 0.00106$, yielding $S_8
\simeq 0.78$. The predicted $S_8$ is within $1\sigma$ of KiDS-1000 measurement,
 $S_8= 0.766^{+0.020}_{-0.014}$~\cite{Heymans:2020gsg}. 

The one-dimensional posterior distribution of the seventh 
parameter $z_c$ is shown in Fig.~\ref{fig:3}. We find that
$\Lambda_s$CDM$^+$ slightly shifts the transition to higher redshifts,
from $z_c = 1.864 \pm 0.149$ to $z_c = 2.105 \pm 0.264$.

\begin{figure}
 \begin{minipage}[t]{0.48\textwidth}
    \postscript{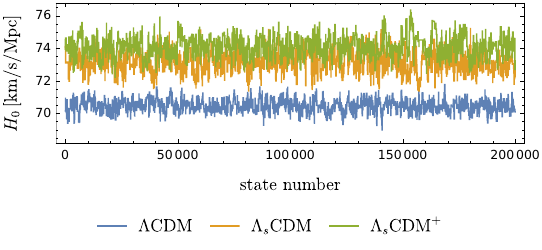}{0.95}
  \end{minipage}
\begin{minipage}[t]{0.48\textwidth}
    \postscript{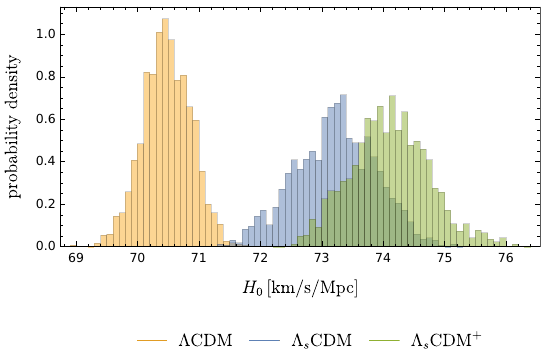}{0.9}
  \end{minipage}
\caption{{\it Top panel.} Markov chains produced by the Metropolis-Hastings algorithm for the
    Hubble constant with the three models under
    consideration. {\it Bottom panel.} Histograms of the states in the
    Markov chains produced by the Metropolis-Hastings algorithm for
    the Hubble constant with three models under
    consideration. \label{fig:4}}
\end{figure}

In Table~\ref{tabla1} we give the minimum values of $\chi^2$ obtained
for each model. One can check by inspection that $\Lambda_s$CDM and
$\Lambda_s$CDM$^+$ are very strongly preferred over
$\Lambda$CDM. The $\chi^2_{\rm min}$ of $\Lambda_s$CDM is smaller than
the one of $\Lambda_s$CDM$^+$ by roughly two units. We can then
conclude that the addition of extra relativistic degrees of freedom
does not spoil the resolution of the cosmic discrepancies presented
in~\cite{Akarsu:2023mfb}. In Fig.~\ref{fig:4} we show the $H_0$ states
of the Markov
chains produced by the three models and the corresponding
histograms. We can see that both $\Lambda_s$CDM and $\Lambda_s$CDM$^+$
exhibit a wider distribution than $\Lambda$CDM.

\section{Conclusions}
\label{sec:4}

$\Lambda_s$CDM is a promising model for solving the $H_0$ and $S_8$
cosmological tensions by demanding the cosmological constant $\Lambda$
to switch sign (from negative to positive) at a critical redshift $z_c \sim 2$~\cite{Akarsu:2019hmw, Akarsu:2021fol, Akarsu:2022typ,Akarsu:2023mfb}. We have reexamined a string-inspired model which can describe the required transition in
  the vacuum energy
making use of the Casimir forces driven by fields
inhabiting the incredible bulk of the dark dimension
scenario~\cite{Anchordoqui:2023woo}. The model, dubbed
  $\Lambda_s$CDM$^+$, deviates from $\Lambda$CDM in the
early universe due to the existence of relativistic neutrino-like
species. We used the Boltzmann solver {\tt CLASS} in combination with
{\tt MontePython} to confront model prediction to experimental
data. We have shown that
$\Lambda_s$CDM$^+$ provides a satisfactory fit to the data and
can resolve the cosmological tensions.

We end by reviewing some key aspects of the $H_0$ conundrum. At the moment, there is a huge set of astrophysical and cosmological observations 
that (on average) seem to show a discrepant trend, in which ``early-time''
model dependent measurements prefer a low $H_0$ ($0.65 < h < 0.70$),
whereas ``late-time'' direct measurements prefer a high $H_0$ ($h\sim 0.73$)~\cite{Abdalla:2022yfr}. To
understand the origin of this discrepancy it is worthwhile to look at
the interplay between $H_0$ and the comoving sound horizon at the end
of the baryonic-drag epoch
$r_{\rm drag}$, which sets the transverse BAO scale. The rationale here is that BAO data play an important role
in defining the $H_0$ tension and also in illuminating possible paths to
resolve it. Let us then focus on current experimental constraints placed on the
$H_0$-$r_{\rm drag}$ plane. Firstly, requiring $\Lambda$CDM
to accommodate Planck data we obtain a  constraint on the sound
horizon $r_{\rm drag} =  147.18 \pm 0.29~{\rm Mpc}$ and
another one on the Hubble constant $h = 0.674 \pm 0.005$~\cite{Planck:2018vyg}. Secondly, the allowed
region of the $H_0$-$r_{\rm drag}$ plane by BAO distance measurements and
uncalibrated supernovae is a U-shaped hyperbolic area; see Fig.~1 in~\cite{Knox:2019rjx}. Actually, the
central values of the allowed region are at the locus of a hyperbola,
because these measurements fix the
product between the Hubble constant and the sound horizon. Finally, the
local $H_0$ measurements by the SHOES team are independent of $r_{\rm drag}$
and thus the allowed region represents a band which is parallel to the
$r_{\rm drag}$ axis. Putting all this together, we can conclude that to resolve
the $H_0$ tension the ``new'' cosmological model should be able to bring
the Planck constraints from $\Lambda$CDM to the region where they can
agree with both {\it (i)} the local SH0ES measurements and {\it (ii)}
the inverse distance ladder measurements from BAO and Hubble flow
supernovae. It is obvious that this requires lowering the sound
horizon, because if the product $H_0 \times r_{\rm drag}$ is fixed,
to increase $H_0$ we must decrease $r_{\rm drag}$. The comoving linear
size of the sound horizon is given by
\begin{equation}
  r_{\rm drag} (z) = \int_{z_{\rm drag}}^\infty \frac{c_s(z')}{H(z')}
  \ dz' \,,
\end{equation}
where
\begin{equation}
  c_s(z) = \frac{c}{\sqrt{3 \{1 + 3 \omega_b/[4 \omega_\gamma (1+z)]\}}}
\end{equation}
 is the sound speed of the photon-baryon fluid and $H(z)$ is the Hubble parameter describing the expansion rate~\cite{Hu:1996vq}. To leading order $c_s \sim c/\sqrt{3}$,
 and so the sound horizon is given by the integral of the inverse of
 the expansion rate. At face value, this implies that to solve the $H_0$ tension we need new physics before the end
of the baryonic-drag epoch, because this is the only way we can reduce the sound horizon.

We can also look at the $H_0$ tension from a perspective which is
independent of  CMB data. To remain totally independent from CMB
measurements, we use data from BAO and uncalibrated supernovae. We take
BAO data calibrated in the early universe (using the sound horizon as a
free parameter) to readjust the absolute value of the expansion rate at
late times, and then we use Hubble flow supernovae to control the shape of
the expansion history. Combining information from these 
observations (while remaining totally independent from the CMB data) we can try reconstructing the late time expansion history of the
Universe. However, in doing so, one inexorably finds that deviations from $\Lambda$CDM are very tightly constrained, 
to no more than 5\% for
$z < 0.6$, which is well below the level required to address the
$H_0$ tension using only new physics at late
times~\cite{Bernal:2016gxb}. This exercise certainly
provides support to our previous finding; namely, that we need new physics before the end
of the baryonic-drag epoch to resolve the $H_0$
tension.\footnote{Arguments have also been given insinuating that
early-time new physics alone is not sufficient to solve the $H_0$
tension~\cite{Vagnozzi:2023nrq}.}

In summary, the $H_0$ tension is actually  a calibrate tension between
the absolute magnitude of supernovae (which controls the local $H_0$
value) and the sound horizon (which controls the absolute scale set by
BAO data). BAO and uncalibrated supernovae alone are very unforgiving,
in the sense that they do not allow huge deviation from $\Lambda$CDM
at late time. All in all, we arrive at a ``no go theorem'' which
states that {\it to resolve the $H_0$ tension we need new physics at
  early times}.

Now, $\Lambda_s$CDM can resolve the cosmological tensions, but does
not deviate from $\Lambda$CDM for $z>z_c$. At first sight this might imply that
$\Lambda_s$CDM violates the $H_0$ ``no go theorem.'' However, we
should remind the reader that in our analysis  we have used the
(angular) transversal 2D BAO data on the shell, which are less model
dependent than the 3D BAO data. This is because the 3D BAO data sample
relies on $\Lambda$CDM to determine the distance to the spherical
shell, and hence could potentially introduce a bias when analyzing
beyond $\Lambda$CDM models~\cite{Bernui:2023byc}. In other words, it
is the 3D BAO data which leave no room for low-$z$ solutions to the
$H_0$ tension~\cite{Gomez-Valent:2023uof}. We can now rephrase the no-go theorem: to address the
$H_0$ tension using the 3D BAO data one needs to consider some sort of new physics at $z > 1000$. If one uses 2D BAO (instead of 3D BAO) data, though, it is possible to solve
the Hubble tension without requiring new physics before
recombination. However, in this case the effective dark energy density
needs to be negative at $z \agt 2$ in order to produce the correct
angular diameter distance to the surface of last scattering.

It is important to
stress that there is an inherent difference between the two models
extending $\Lambda$CDM: in
contrast to $\Lambda_s$CDM, the proposed $\Lambda_s$CDM$^+$ carries
new physics in the early universe, which modifies the value of $r_{\rm drag}$ predicted by $\Lambda$CDM.

In closing we note that $\Lambda_s$CDM has a hidden sudden singularity
at $z_c$~\cite{Barrow:2004xh}. Indeed, it is easily seen that the
scale factor $a$ is continuous and non-zero at $t=t_c$, but its first
derivative $\dot a$ is discontinuous, and its second derivative
$\ddot a$ diverges.\footnote{It has been shown  that the sudden
  singularity at $z_c$ 
does not result in the dissociation of bound
systems~\cite{Paraskevas:2024ytz}.} To a first approximation, we have
implemented the model into the {\tt CLASS} Boltzmann solver using a signum function to
characterize the tunneling of $\varphi$. In the spirit
of~\cite{Akarsu:2024qsi}, an investigation to smoothed-out this sharp
transition is obviously important to be done.\\

\section*{Acknowledgements}
The work of L.A.A. is supported by the U.S. National
Science Foundation (NSF Grant PHY-2412679). I.A. is supported by the Second Century Fund (C2F), Chulalongkorn University. The work of D.L. is supported by the Origins
Excellence Cluster and by the German-Israel-Project (DIP) on
Holography and the Swampland. N.T.N. is supported by the AstroCom NYC program through NSF Grant AST-2219090.

\section*{Appendix: Orientifolding and $\bm{T_4}$}

Anomaly and tadpole cancellation conditions have long been used as
intriguingly related constraints in the construction of consistent
vacuum configurations with open and unoriented strings~\cite{Polchinski:1987tu,Polchinski:1995mt,Gimon:1996rq,Bianchi:2000de,Angelantonj:2002ct}. In particular, the tadpole
constraint implies that the total Ramond-Ramond (RR) 3-brane charge supported by the compact manifold must vanish; 
\begin{equation}
  N_{\rm flux} + N_{{\rm D}3} = \frac{1}{4} N_{{\rm O}3} \,, 
\label{AppenEq}
\end{equation}
where $N_{\rm flux}$ is the 3-form flux which is positive,  $N_{{\rm D}3}$
is the number of 3D-branes with a positive RR charge (normalized to 1), and
$N_{{\rm O}3}$ is the number of orientifolds with a negative RR charge of $-1/4$~\cite{Cascales:2003zp}. If the flux is
supersymmetric, the tension satisfies the same relation
(dimensionalities are taken care by the  inverse string tension $\alpha'$). This implies that the sum of positive tensions from branes and fluxes should cancel the negative tensions
  from orientifolds, so the total tension vanishes. However, this is a
  global condition which is  not true locally. The SM could be located together with orientifolds
  (indeed orientifolds and D-branes can be on top of each other) so
  that locally the tension is negative; the Randall-Sundrum model~\cite{Randall:1999ee} is a
  textbook example. Note that although the total
  localize tension could be negative, the SM lives on
a D-brane (and not on an orientifold or a flux). In non-supersymmetric constructions, Eq.~(\ref{AppenEq}) is always satisfied for the charge, but not necessarily
for the tension since the objects do not preserve the same linearly
realized supersymmetry. This introduces an additional freedom in model
building to accommodate a negative $T_4$~\cite{Dudas:2004nd}.

\end{document}